\documentclass[usenatbib,usegraphicx]{mn2e}
\usepackage{amssymb}
\usepackage{graphicx}
\usepackage{mathrsfs}
\usepackage{longtable}

\newcommand{\OVIdblt}{{O}\kern 0.1em{\sc vi}~$\lambda\lambda 1032, 1038$}
\newcommand{\CII}{\hbox{{C}\kern 0.1em{\sc ii}}}
\newcommand{\CIII}{\hbox{{C}\kern 0.1em{\sc iii}}}
\newcommand{\CIV}{\hbox{{C}\kern 0.1em{\sc iv}}}
\newcommand{\HI}{\hbox{{H}\kern 0.1em{\sc i}}}
\newcommand{\Lya}{\hbox{{Ly}\kern 0.1em$\alpha$}}
\newcommand{\Lyb}{\hbox{{Ly}\kern 0.1em$\beta$}}
\newcommand{\Lyg}{\hbox{{Ly}\kern 0.1em$\gamma$}}
\newcommand{\Lyd}{\hbox{{Ly}\kern 0.1em$\delta$}}
\newcommand{\Lye}{\hbox{{Ly}\kern 0.1em$\epsilon$}}
\newcommand{\Lyz}{\hbox{{Ly}\kern 0.1em$\zeta$}}
\newcommand{\Lyeta}{\hbox{{Ly}\kern 0.1em$\eta$}}
\newcommand{\MgII}{\hbox{{Mg}\kern 0.1em{\sc ii}}}
\newcommand{\OVI}{\hbox{{O}\kern 0.1em{\sc vi}}}
\newcommand{\OVII}{\hbox{{O}\kern 0.1em{\sc vii}}}
\newcommand{\OVIII}{\hbox{{O}\kern 0.1em{\sc viii}}}
\newcommand{\NV}{\hbox{{N}\kern 0.1em{\sc v}}}
\newcommand{\SiII}{\hbox{{Si}\kern 0.1em{\sc ii}}}
\newcommand{\SiIII}{\hbox{{Si}\kern 0.1em{\sc iii}}}
\newcommand{\SiIV}{\hbox{{Si}\kern 0.1em{\sc iv}}}
\newcommand{\FeII}{\hbox{{Fe}\kern 0.1em{\sc ii}}}

\newcommand{\NeIX}{\hbox{{Ne}\kern 0.1em{\sc ix}}}
\newcommand{\OI}{\hbox{{O}\kern 0.1em{\sc i}}}

\def\apj{ApJ}

\title[Radiation-Driven Galactic Winds]
{Radiation Pressure Driven Galactic Winds from Self-Gravitating Discs}
\author[Zhang \& Thompson]
{Dong Zhang$^{1}$\thanks{E-mail:dzhang@astronomy.ohio-state.edu}
and Todd A.~Thompson$^{1,2}$\thanks{Alfred P.~Sloan Fellow}\\
$^1$Department of Astronomy, The Ohio State University, 140 West 18th Avenue, Columbus, OH 43210, USA\\
$^2$Center for Cosmology \& Astro-Particle Physics, The Ohio State University, Columbus, Ohio 43210, USA}

\begin{document}

\maketitle

\begin{abstract}

We study large-scale winds driven from uniformly bright self-gravitating
discs radiating near the Eddington limit. We show that the ratio of
the radiation pressure force to the gravitational force increases
with height above the disc surface to a maximum of twice the value of the ratio at the disc surface.
Thus, uniformly bright self-gravitating discs radiating at the
Eddington limit are fundamentally unstable to driving large-scale
winds. These results contrast with the spherically symmetric case,
where super-Eddington luminosities are required for wind formation.
We apply this theory to galactic winds from rapidly star-forming galaxies that
approach the Eddington limit for dust. For hydrodynamically coupled
gas and dust, we find that the asymptotic velocity of the wind is
$v_{\infty}\simeq1.5\,v_{\rm rot}$ and that
$v_{\infty}\propto {\rm SFR}^{0.36}$, where $v_{\rm
rot}$ is the disc rotation velocity and SFR is the star
formation rate, both of which are in agreement with observations.
However, these results of the model neglect the gravitational
potential of the surrounding dark matter halo and a (potentially
massive) old passive stellar bulge or extended disc, which act to decrease $v_\infty$. A more
realistic treatment shows that the flow can either be unbound, or
bound, forming a ``fountain flow'' with a typical turning timescale
of $t_{\rm turn}\sim0.1-1$ Gyr, depending on the ratio of the mass
and radius of the rapidly star-forming galactic disc relative to the total mass and
break (or scale) radius of the dark matter halo or bulge. We provide
quantitative criteria and scaling relations for assessing whether or
not a rapidly star-forming galaxy of given properties can drive unbound flows via the
mechanism described in this paper. Importantly, we note that because
$t_{\rm turn}$ is longer than the star formation timescale (gas
mass/star formation rate) in the rapidly star-forming galaxies and ultra-luminous
infrared galaxies for which our theory is most applicable, if
rapidly star-forming galaxies are selected as such, they may be observed to have strong
outflows along the line of sight with a maximum velocity $v_{\rm max}$
comparable to $\sim1.5\,v_{\rm rot}$, even
though their winds are eventually bound on large scales.

\end{abstract}


\section{Introduction}
Galactic-scale winds are ubiquitous in rapidly star-forming galaxies in both
the local and high-redshift universe (\citealt{Heckman90};
\citealt{Heckman00}; \citealt{Pettini01, Pettini02};
\citealt{Shapley03}; \citealt{rupke05}; \citealt{Sawi08}). They are
important for determining the chemical evolution of galaxies and the
mass-metallicity relation (\citealt{DS86}; \citealt{Tremonti04};
\citealt{Erb06}; \citealt{finlator}; \citealt{Peeples11}), and as a
primary source of metals in the intergalactic medium (IGM; e.g.,
\citealt{aguirre01}). Moreover, galactic winds are perhaps the most
extreme manifestation of the feedback between star formation in a
galaxy and its interstellar medium (ISM). This feedback mechanism is
crucial for understanding galaxy formation and evolution over cosmic
time (\citealt{SH03}; \citealt{OD06}; \citealt{OD08};
\citealt{Oppen10}).

The most well-developed  model for galactic winds from rapidly star-forming galaxies is
the supernova-driven model of \cite{CC85}, which assumes that the
energy from multiple stellar winds and core-collapse supernovae in
the starburst is efficiently thermalized.  The resulting hot flow
drives gas out of the host, sweeping up the cool ISM
(\citealt{Heckman93}; \citealt{DeYoung94}; \citealt{SS00}; \citealt{strickland02};
\citealt{SH09}; \citealt{Fujita09}). Although this model is
successful in explaining the X-ray properties of rapidly star-forming galaxies, the
recent observational results that galaxies with higher star
formation rates (SFRs) accelerate the absorbing cold gas clouds to
higher velocities ($v\propto$ SFR$^{0.35}$) and that the wind
velocity is correlated with the galaxy escape velocity may challenge
the traditional hypothesis that the cool gas is accelerated by the
ram pressure of the hot supernova-heated wind, whose X-ray emission
temperature varies little with SFR, circular velocity, and host
galaxy mass, indicating a critical galaxy mass below which most of
the hot wind escapes (e.g., \citealt{Martin99}). These new observations may instead favor
momentum-driven or radiation pressure-driven models for the wind
physics (e.g., \citealt{Martin05}; \citealt{Weiner09}; Murray, Quataert \& Thompson 2005 [hereafter
MQT05]).

The model that galactic winds may be driven by momentum deposition
provided by radiation pressure from the continuum absorption and
scattering of starlight on dust grains was developed by MQT05. However, the
conclusions of MQT05 are based on an assumed isothermal potential
and spherical geometry, and are thus most appropriate for
bright elliptical/spheroidal galaxies in formation. On the other hand, the theory of
radiation-driven winds from accretion discs from the stellar to
galactic scales has also been studied (e.g., \citealt{TF96, TF98};
\citealt{Proga98, Proga99}; \citealt{Proga00, Proga03}), but none of
these works considered radiation from self-gravitating discs.

In this paper we answer the question of whether or not large-scale
winds can be driven by radiation pressure from self-gravitating
discs radiating near the Eddington limit. These considerations are
motivated by the work of Thompson, Quataert \& Murray (2005; TQM05),
who argued that radiation pressure on dust is the dominant feedback
mechanism in rapidly star-forming galaxies, and that in these systems star formation is
Eddington-limited. For simplicity, throughout this paper we assume
that the disc is of uniform brightness and surface density. In Section 2,
we show that such discs are fundamentally unstable to wind formation
because the radiation pressure force dominates gravity in the
vertical direction above the disc surface. This result is
qualitatively different from the well-known case in spherical
symmetry. In Section 2, we also discuss the applicability of this model to
rapidly star-forming galaxies and then calculate the terminal velocity of the wind
along the disc pole, and its dependence on both the SFR and galaxy
escape velocity. In Section 3, we assess the importance of a spherical
stellar bulge or extended passive disc, and dark matter halo potential. In Section 4, we discuss the
3-dimensional wind structure and estimate the total wind mass loss
rate. We discuss our findings and conclude in Section 5.

\section{Radiation-Driven Winds \& The Terminal Velocity}\label{SimpleModel}
We consider the idealized model problem of a disc with uniform brightness and total surface
density: $I(r\leq r_{\rm rad})=I$ and $\Sigma(r\leq r_D)=\Sigma$,
where $r_{\rm rad}$ and $r_D$ define the outer radius of the
luminous and gravitating portion of the disc, respectively. The
flux-mean opacity to absorption and scattering of photons is
$\kappa$. The gravitational force along the polar axis above the
disc is
\begin{eqnarray}
f_{\rm grav}(z)&=&-2\pi G\Sigma\int_{0}^{r_{D}}\frac{zrdr}{(r^{2}+z^{2})^{3/2}}\nonumber\\
&=&-2\pi G\Sigma\left(1-\frac{z}{\sqrt{z^{2}+r_{D}^{2}}}\right),\label{integral01}
\end{eqnarray}
and the vertical radiation force along the pole is
\begin{equation}
f_{\rm rad}(z)=\frac{2\pi \kappa I}{c}\int_{0}^{r_{\rm rad}}\frac{z^{2}rdr}{(r^{2}+z^{2})^{2}}
=\frac{\pi\kappa I}{c}\frac{r_{\rm rad}^{2}}{z^{2}+r_{\rm rad}^{2}}.\label{integral02}
\end{equation}
The extra factor of $z/\sqrt{r^{2}+z^{2}}$ in the radiation force
integral compared with the gravitational force is the projection cosine factor $\cos \theta$ such that $df_{\rm rad}\propto\cos \theta\,d\Omega$, where $\theta$ is the angle between the $z$-axis direction and the direction of the solid angle $d\Omega$ at the disk surface (\citealt{Rybicki79}; see also \citealt{Proga98}; \citealt{TF98}). Thus, the Eddington
ratio along the pole $\Gamma(z)=|f_{\rm rad}(z)/f_{\rm grav}(z)|$
as a function of height $z$ is given by
\begin{equation}
\Gamma(z)=\Gamma_{0}\left(\frac{r_{\rm rad}}{r_{D}}\right)^{2}
\left(\frac{z^{2}+r_{D}^{2}}{z^{2}+r_{\rm rad}^{2}}+\frac{z\sqrt{z^{2}+r_{D}^{2}}}{z^{2}+r_{\rm rad}^{2}}\right)\label{ratio},
\end{equation}
where $\Gamma_{0}=\Gamma(z=0)=\kappa I/(2cG\Sigma)$ is the Eddington
ratio at the disc center. A disc at the Eddington limit
($\Gamma_{0}=1$) requires $I_{\rm Edd}=2cG\Sigma/\kappa$, or flux
$F_{\rm Edd}=2\pi cG\Sigma/\kappa$. If $r_{\rm rad}/r_{D}>1/\sqrt{2}\simeq0.7$, then $\Gamma(z)$ increases along
the $z$-axis above the disc. In particular, for $r_{\rm rad}\simeq
r_{D}$, the radiation force becomes twice the gravitational force as
$z\rightarrow\infty$:
\begin{equation}
\Gamma_{\infty}=\Gamma(z\rightarrow\infty)=2.
\end{equation}
Because $\Gamma(z)$ increases monotonically with $z$, an
infinitesimal displacement of a test particle in the vertical
direction yields a net vertical acceleration, and the disc is thus
unstable to wind formation. This result for discs is qualitatively
different from the spherical case with a central point source where
$\Gamma=f_{\rm rad}/f_{\rm grav}$ is constant with radius.

In the more realistic case we consider an exponential disc with surface brightness $I=I_{0}\exp(-r/R_{\rm rad})$
and surface density $\Sigma=\Sigma_{0}\exp(-r/R_{D})$. The disc has approximately uniform
brightness for $r<R_{D}$ and uniform density for $r<R_{\rm rad}$, while the brightness and mass distribution cut
off for $r\simeq R_{\rm rad}$ and $r\simeq R_{D}$ respectively. The vertical radiation and gravitational forces
along the pole above the exponential disc can be calculated by multiplying
the integrals in equations (\ref{integral01}) and (\ref{integral02})
by the exponential terms $\exp(-r/R_{D})$ and $\exp(-r/R_{\rm rad})$ respectively.
As a result, in this case the Eddington ratio becomes a function of $R_{\rm rad}/R_{D}$.
We obtain the Eddington ratio along the pole $\Gamma(z\rightarrow\infty)=2\Gamma(z=0)$
if $R_{\rm rad}=R_{D}$, and $\Gamma(z\rightarrow\infty)\leq\Gamma(z=0)$ if $R_{D}\geq\sqrt{2}R_{\rm rad}$.
Therefore the exponential disc gives a result similar to the uniform disc
case in that $\Gamma(z)$ increases monotonically only if
$R_{D}<\sqrt{2}R_{\rm rad}\simeq1.4R_{\rm rad}$, otherwise the disc is not promising to drive a wind because the gravitational force dominates
over the outward radiation pressure force. Then we take $r_{D}=\sqrt{2}R_{\rm rad}$ and refer to the region outside of the active and bright region ($r \gtrsim \sqrt{2} R_{\rm rad}$) as the ``extended passive disc'', and we discuss it together with the effects of an old stellar bulge and dark matter halo in Section \ref{BulgeHalo}.

\begin{figure}
\centerline{\includegraphics[width=9cm]{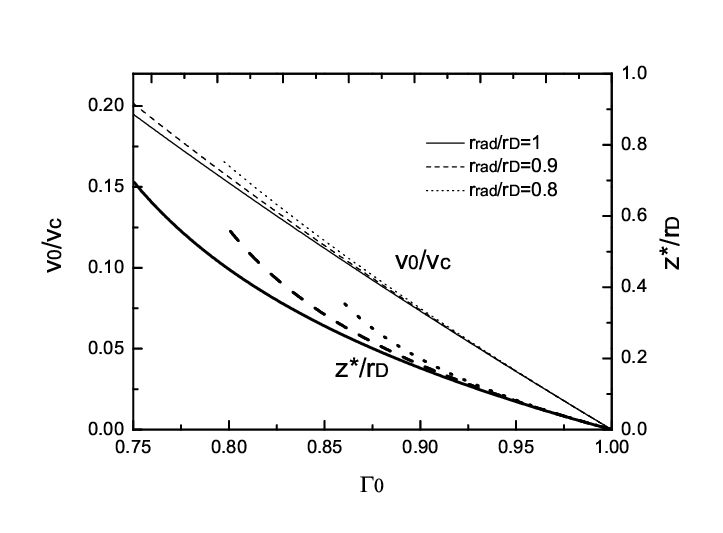}}
\caption{Minimum initial velocity $v_{0}/v_{c}$ to drive unbound winds (thin lines),
and the critical height $z^{*}$ turning the gas from deceleration to accelerations (thick lines)
as a function of Eddington ratio $\Gamma_{0}$ in the sub-Eddington case $2/(\pi\hat{r})<\Gamma_{0}<1$,
where $\hat{r}=r_{\rm rad}/r_{D}$=1, 0.9 and 0.8.}\label{fig_subEdd}
\end{figure}

Hereafter we adopt the uniform disc model. If we consider the motion of a test particle in the outflow, the
velocity along the $z$-axis can be written as
\begin{eqnarray}
\frac{v(z)^{2}-v_{0}^{2}}{4\pi G\Sigma r_{D}}=\hat{r}\Gamma_{0}\arctan\left(\frac{\hat{z}}{\hat{r}}\right)
-\left(1+\hat{z}-\sqrt{1+\hat{z}^{2}}\right),\label{orbit01}
\end{eqnarray}
where $\hat{r}=r_{\rm rad}/r_{D}$, $\hat{z}=z/r_{D}$, and $v_{0}$ is
initial vertical velocity. The first term on the right side of
equation (\ref{orbit01}) is the ``radiation potential'' along the
pole, while the second term is the gravitational potential. The
right side is always positive if $\Gamma_{0}\geq1$ and
$\hat{r}\Gamma_{0}>2/\pi\approx0.64$, and thus the gas can be
accelerated to infinity. On the other hand, an unbound outflow is
still possible in the sub-Eddington case $\Gamma_{0}<1$ if the
initial velocity $v_{0}$ is sufficiently large to escape the
gravitational potential above the disc until reaching the critical
height $z^{*}$ where $\Gamma(z^{*})=1$, because the gas is
decelerated from its initial $v_{0}$ until it reaches $z^{*}$, it
will be accelerated to infinity if $v(z^{*})\geq0$. Using this
constraint, we calculate the minimum $v_{0}$ required to drive an
unbound wind in the sub-Eddington case, and the critical height
$z^{*}$ where the wind profile acceleration changes sign. Figure
\ref{fig_subEdd} shows the results. We introduce a characteristic
velocity $v_{c}=\sqrt{4\pi G\Sigma r_{D}}$. The calculation is
applied for $v_{\infty}=v(z\rightarrow\infty)\geq v_{0}$, or
$\hat{r}\Gamma_{0}>2/\pi\simeq0.64$. In this case the wind can be
accelerated at infinity. The important point here is that
for a disc with sub-Eddington brightness, and non-zero vertical velocity $v_0$,
the gas can be first
decelerated and then accelerated to infinity.
For the typical random initial velocities of gas in galaxies $\rho
v_{0}^{2}\sim\pi G \Sigma^{2}$, we have
$v_{0}/v_{c}\sim(h/2r_{D})^{1/2}\sim0.22(r/r_{D})^{1/2}$, where the
galactic thickness scale $h=0.1r$ has been assumed for the second
equality (\citealt{1998ApJ...507..615D}). For this reason we expect
that even somewhat sub-Eddington discs can drive outflows, as shown in Figure
\ref{fig_subEdd}. In contrast with the spherical case for which
$v_{z}^{2}-v_{0}^{2}\propto \Gamma_{0}-1$, uniformly bright
self-gravitating discs allow the gas above the disc to be
accelerated for the case with $\Gamma_{0}=1$ or even for some
sub-Eddington cases. In Section 4, we show the 3-dimensional
trajectories of wind particles launched from both Eddington
and sub-Eddington discs.

We wish to apply this simple theory to rapidly star-forming galaxies, which may
reach the Eddington limit for dust (TQM05).  However, this
application depends on the extent to which a disc-like collection
of point sources of radiation (stars) may be treated as a uniformly
bright disc. In the limit that the dusty gas of rapidly star-forming galaxies is
optically-thick to the re-radiated FIR emission in galactic discs
($\Sigma\gtrsim0.1-1$\,g cm$^{-2}$; $\kappa_{\rm
FIR}\sim1-10$\,cm$^2$ g$^{-1}$; see TQM05), as is reached in ULIRGs,
the self-gravitating discs around bright AGN, and some rapidly star-forming galaxies,
the approximation of a uniform brightness disc is likely valid.
However, for $\Sigma\lesssim10^{-3}$\,g cm$^{-2}$ ($\kappa_{\rm
UV}\sim10^3$\,cm$^2$ g$^{-1}$), the disc is optically thin to the UV
radiation of massive stars. Because the massive stars are point sources and dominate
the galaxy's bolometric luminosity, the approximation of a 2-dimensional
uniformly bright disc breaks down. In the intermediate regime
$10^{-3}\lesssim\Sigma\lesssim0.1-1$\,g cm$^{-2}$,\footnote{ These
bounds on $\Sigma$ depend linearly on the gas-to-dust ratio.} the
disc is optically-thick to UV radiation, but optically-thin to the
re-radiated FIR, the application of the simple uniformly bright disc
model is tentative, as the fraction of the light that is absorbed and
scattered in the gas depends on the distribution of the sources
relative to the dusty gas, the inter-star spacing relative to the
vertical scale of the dusty gas, and the grain size distribution,
which affects the overall scattering albedo. We save a detailed
investigation of the intermediate case for a future work, but here note two important points:
(1) even for low-$\Sigma$ galaxies, the fraction of diffuse
(potentially scattered) UV light may be substantial
(\citealt{Thilker05}), and (2) if a finite thickness disc
is uniformly emissive and absorptive, then $\kappa_{\rm UV}F_{\rm UV}\sim\kappa_{\rm UV}F_{\rm tot}/\tau_{\rm
UV}\sim F_{\rm tot}/\Sigma>\kappa_{\rm FIR}F_{\rm FIR}$, where
$\tau_{\rm UV}\sim \kappa_{\rm UV}\Sigma$ and $F_{\rm tot}$, $F_{\rm UV}$,
and $F_{\rm FIR}$ are the total, UV, and FIR fluxes, respectively.
The radiation pressure force above the disc is thus dominated by UV emission from
the ``skin'' of the disc (the $\tau_{\rm UV}\sim1$ surface). Therefore, if the total flux
is equal to the Eddington flux -- $F_{\rm tot}=2\pi Gc\Sigma/\kappa$, where
$\kappa$ is the flux-mean opacity -- then the escaping UV radiation is also at Eddington:
$F_{\rm UV}\sim F_{\rm tot}/\tau_{\rm UV}\sim2\pi Gc\Sigma/(\tau_{\rm UV}\kappa)\sim 2\pi Gc\Sigma/\kappa_{\rm UV}$.
For these reasons, the model of a uniformly bright disc may apply even when
$\Sigma<0.1-1$ g cm$^{-2}$ and the medium is not optically thick to the reradiated FIR.


To proceed with our application to rapidly star-forming galaxies, the characteristic
velocity $v_{c}$ is written as
\begin{equation}
v_{c}=\sqrt{4\pi G\Sigma r_{D}}=500\,\textrm{km}\,\textrm{s}^{-1}\Sigma_{0}^{1/2}r_{D,1kpc}^{1/2},\label{terminal00}
\end{equation}
where we take $\Sigma_{0}=\Sigma/1$ g cm$^{-2}=\Sigma/(4800\,{\rm
M}_{\odot}$ pc$^{-2}$), and $r_{D,1kpc}=r_{D}/1$ kpc.
Momentarily neglecting the importance of the surrounding
dark matter halo, or a potentially massive old stellar
bulge or extended passive disc (which we evaluate in Section 3),
from equation
(\ref{orbit01}) with $\Gamma_0=1$ and $r_{\rm rad}\simeq r_{D}$, the
asymptotic terminal velocity along the pole is
\begin{equation}
v_{\infty}=v_{c}\sqrt{\pi/2-1}\simeq380\,\textrm{km}\,\textrm{s}^{-1}\Sigma_{0}^{1/2}r_{D,1kpc}^{1/2}.\label{terminal01}
\end{equation}
This expression for $v_\infty$ can be related to the star formation
rate (SFR) using the Schmidt law, which relates the star formation
surface density and gas surface density in galactic discs:
$\Sigma_{\rm SFR}\propto\Sigma_{\rm gas}^{1.4}$ (\citealt{Ken98}).
We approximate $\Sigma_{\rm gas}=0.5f_{g,0.5}\Sigma$, where $f_{g}$
is the gas fraction. Since SFR$\simeq\Sigma_{\rm SFR}\pi r_{D}^{2}$,
we have $v_{\infty}\propto\Sigma^{1/2}r_{D}^{1/2}\propto\Sigma_{\rm
gas}^{1/2}r_{D}^{1/2}\propto\Sigma_{\rm SFR}^{0.36}r_{D}^{1/2}\propto(\textrm{SFR}/r_{D}^{2})^{0.36}r_{D}^{1/2}\propto\textrm{SFR}^{0.36}r_{D}^{-0.21}$ or
\begin{eqnarray}
v_{\infty}&\sim&150\,\textrm{km}\,\textrm{s}^{-1}f_{g,0.5}^{-0.5}\left(\frac{\Sigma_{\rm SFR}}{M_{\odot}\,\textrm{yr}^{-1}\,\textrm{kpc}^{-2}}\right)^{0.36}
r_{D,1kpc}^{0.5}\nonumber\\
&\sim&400\,\textrm{km}\,\textrm{s}^{-1}f_{g,0.5}^{-0.5}\left(\frac{\rm SFR}{50M_{\odot}\textrm{yr}^{-1}}\right)^{0.36}
r_{D,1kpc}^{-0.21},\label{terminal02}
\end{eqnarray}
which is consistent with the observation $v_{\infty}\propto$
SFR$^{0.35\pm0.06}$ in low-redshift ULIRGs (\citealt{Martin05}) and
$v_{\infty}\propto$ SFR$^{0.3}$ for high-stellar-mass and high-SFR
galaxies at redshift $z\sim1$ (\citealt{Weiner09}; but, see Fig.~17
from \citealt{Chen10}). The observed scatter at a given SFR may be
caused by different $r_{D}$, $f_g$, $r_{\rm rad}/r_{D}$, $v_{0}$, bulge and
dark matter halo mass, and the time dependence of the wind properties as the stellar population evolves (see Section 3).
Since we only use a simplified uniform disc model to derive equations
(\ref{terminal01}) and (\ref{terminal02}), a more definitive
comparison with the data should await a model with more realistic
distributions of surface density, opacity, and brightness.
Moreover, we can give the criterion for driving a galactic wind in terms of SFR or $\Sigma_{\rm SFR}$.
For an Eddington-limited disc, we have $F_{\rm Edd}\propto\Sigma_{\rm gas}\propto\Sigma_{\rm SFR}^{0.71}\propto($SFR$/r_{D}^{2})^{0.71}\propto$ SFR$^{0.71}r_{D}^{-1.43}$ or
\begin{eqnarray}
F_{\rm Edd}&\simeq&5\times10^{11}f_{g,0.5}^{-1}\kappa_{1}^{-1}\left(\frac{\Sigma_{\rm SFR}}{M_{\odot}\,\textrm{yr}^{-1}\,\textrm{kpc}^{-2}}\right)^{0.71}\;L_{\odot}\;\textrm{kpc}^{-2}\nonumber\\
&\simeq&3\times10^{12}f_{g,0.5}^{-1}\kappa_{1}^{-1}r_{D,1kpc}^{-1.43}
\left(\frac{\rm SFR}{50M_{\odot}\textrm{yr}^{-1}}\right)^{0.71}\;L_{\odot}\;\textrm{kpc}^{-2},\label{Fedd}
\end{eqnarray}
where we have again employed the Schmidt law and $\kappa=10\,\kappa_{1}$ cm$^{2}$ g$^{-1}$ is the flux-mean dust opacity.
Keep in mind that the dust opacity in the FIR limit is about $\kappa_{\rm FIR}\sim1-10$ cm$^{2}$ g$^{-1}$.

If we assume the galactic disc is in radial centrifugal balance with
a Keplerian velocity $v_{\rm rot}\sim\sqrt{G\pi\Sigma r}$, and we
take the typical terminal velocity from equation (\ref{terminal01})
to compare to the rotation velocity at the radius $r_{D}$ that $v_{\rm rot}=v_{\rm rot}(r=r_{D})$, we have
\begin{equation}
v_{\infty}\simeq2\sqrt{\pi/2-1}\,v_{\rm rot}
\simeq1.5\,v_{\rm rot}\label{velocity_1}.
\end{equation}
For an exponential disc with brightness $I\propto\exp(-r/R_{\rm rad})$, the peak rotation velocity is
$v_{\rm rot}\simeq0.8\sqrt{G\pi \Sigma_{0} R_{\rm rad}}$, which is approximately reached at $r\simeq 1.8 R_{\rm rad}$. This estimate is again made in the absence of galactic bulge or extended passive disc and dark matter halo. That the terminal velocity of the wind increases linearly
with the galactic rotation velocity is also consistent with the
observational results in Martin (2005, Fig 7). Because the flat part of the rotation curve $v_{c}$ is
typically comparable to the rotation velocity at the edge of the galactic disk (\citealt{Sharma11}), thus we also have $v_{\infty} \sim 1.5\,v_{c}$.
However, the factor ``1.5'' in equation (\ref{velocity_1}) cannot be strictly obtained due to the gravitational potential of the dark matter halo or a spherical old stellar bulge or extended passive disc, which decrease the factor of 1.5 and can cause the flow to become bound (see Section 3).

In their cosmological simulations of structure formation
and IGM enrichment by galactic winds, \cite{OD06} assumed a wind launch
velocity $v_{\rm launch}=3\sigma\sqrt{\Gamma_0-1}\sim3\sigma$, where $\sigma$
is the galactic velocity dispersion and the factor of 3 is an
assumption (see also \citealt{OD08}; \citealt{Oppen10}).
Taking the relation between velocity dispersion and rotation velocity $\sigma=v_{\rm rot}/\sqrt{2}$ for the isothermal sphere (\citealt{Binney08}),
the wind launch velocity can be written as $v_{\rm launch}\sim2\,v_{\rm rot}$, similar to equation (\ref{velocity_1}). However, besides the initial wind launch velocity, an additional kick is given to the wind particles
to overcome the halo potential, implying that the scaling of wind velocity $v_{\infty}$
in Oppenheimer \& Dav\'{e}'s model has a higher amplitude than what we obtain from
just the effect of radiation pressure on dust discussed in this paper. Nevertheless, equations
(\ref{terminal01}) - (\ref{velocity_1})
are essential for developing an understanding of the maximum
envelope of values for $v_\infty$, and its dependence on the
observed properties of rapidly star-forming galaxies.
More physics of $v_{\infty}$ and its z-dependence in the
gravitational potential well of a spherical old stellar bulge
or dark matter halo is presented in Section 3.

Finally, we note that the characteristic timescale for the wind to
reach its asymptotic velocity is
\begin{equation}
t_{c}\sim\sqrt{r_D/(4G\Sigma)}\sim3.5\times10^{6}\Sigma_{0}^{-1/2}r_{D,1kpc}^{1/2}\,\,\textrm{yr},
\label{tc}
\end{equation}
which can be compared with the timescale of a bright star-forming
disc:  $t_{\star}\sim M_{\rm gas}/$SFR. Again employing the Schmidt
Law, we find that
\begin{equation}
t_\star\sim2\times10^{8}f_{g,0.5}^{-0.4}\Sigma_{0}^{-0.4}\,\,\,{\rm yr}
\label{tstar}
\end{equation}
or $t_c/t_\star\sim0.02
r_{D,1kpc}^{1/2}f_{g,0.5}^{0.4}\Sigma_0^{-0.1}$. Since $t_{c}<t_{\star}$, the wind
can be accelerated to $\sim v_\infty$ with a timescale of $t_{c}$ before the gas supply is
depleted by star formation in a timescale of $t_{\star}$. Note that gas recycling in the ISM will give an even longer $t_\star$ than the simple estimate $M_{\rm gas}$/SFR.


\section{Extended Disc, Bulge and Dark Matter Halo}\label{BulgeHalo}

\begin{figure*}
\centerline{\includegraphics[width=18cm]{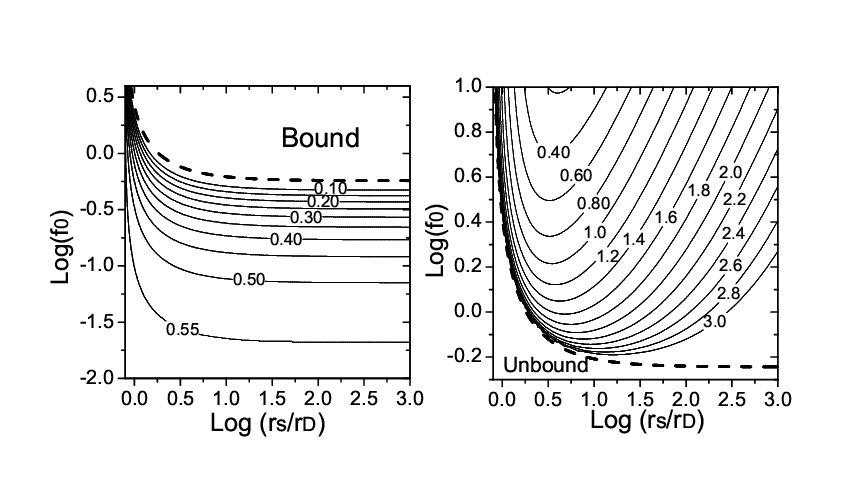}}
\caption{Contours of the asymptotic kinetic energy
$\left(v_{\infty}/v_{c}\right)^{2}$ as a function of $f_0=f_b+f_h$
(assuming $f_b=0$) and $c'=r_s/r_D$ for unbound particles
(\textit{left}), and the turning point $\log_{10}(z_{\rm
turn}/r_{D})$ along the pole in the case of bound particles,
assuming $v_{0}=0$ (\textit{right}).  The dashed line in both panels
divides ``bound'' and ``unbound'' trajectories.}\label{fig_halo}
\end{figure*}

The galactic bulge, extended passive stellar or gaseous disc and dark matter halo are important to the wind dynamics. If we do not consider the luminosity from the galactic bulge and extended disc,\footnote{The case of a bright spherical
bulge was considered in MQT05.} the bulge, extended disc and halo only act
to decrease the asymptotic wind velocity, and may cause the
wind to fall back to the disc as a ``fountain flow''. For the galactic
bulge we employ a truncated constant density sphere of mass $M_{\rm bulge}$ and spherical radius $r_{\rm bulge}$. The effect of an old passive extended stellar or disc of gas can be neglected within a height scale $z \lesssim (\Sigma/\Sigma_{\rm ext})^{1/2}R_{D}$ with $\Sigma$ and $\Sigma_{\rm ext}$ being the surface density of nuclear disc ($r<\sqrt{2}R_{\rm rad}$) and extended star/gas disc ($r>\sqrt{2}R_{\rm rad}$). On the other hand, its effect at large scales is similar to the effect of an old stellar bulge, but with the following replacements: $M_{\rm bulge}\rightarrow M_{\rm ext}$ and $r_{\rm bulge}\rightarrow R$, where $R=\sqrt{r^{2}+z^{2}}$ is the cylindrical distance to the halo center, and $M_{\rm ext}$ is the total mass of extended disc. Note that for normal spirals and field galaxies in the local universe, their extended star/gas discs can be very large compared to the region of more rapid star formation. As discussed in Section \ref{SimpleModel}, where we identified the requirement $R_{D}<\sqrt{2} R_{\rm rad}$ for wind formation, they are thus not promising to drive galactic winds via the mechanism discussed here. Instead, bulgeless galaxies without extended discs, and with very massive central concentrations of star formation --- e.g., those ULIRGs with very large nuclear surface density and star formation rate, but with small $\Sigma_{\rm ext}$--- are most promising.
For simplicity we do not consider the importance of an passive disc further here. Also, we adopt
the NFW potential (\citealt{NFW96}) to describe the dark matter halo
distribution $\rho_{\rm DM}(r)\propto R^{-1}(R+r_{s})^{-3}$, where
$r_{s}$ is the scale radius. For simplicity, in this section we take
the uniform disc model with $r_{\rm rad}= r_{D}$.
The Eddington limit $\Gamma_{0}=1$ including the dark matter halo becomes
\begin{equation}
\frac{\pi\kappa I}{c}=2\pi G\Sigma+\frac{1}{2}\frac{GM_{\rm halo}}{r_{s}^{2}f(c_{\rm vir})}=2\pi G\Sigma\left(1+\frac{f_{h}}{2c'}\right),
\end{equation}
where $c_{\rm vir}=r_{\rm vir}/r_{s}$, $r_{\rm vir}$ is the virial radius,
$f(c_{\rm vir})=\ln(1+c_{\rm vir})-c_{\rm vir}/(1+c_{\rm vir})$, and
\begin{equation}
f_{h}=\frac{M_{\rm halo}}{2\pi r_{s}r_{D}\Sigma f(c_{\rm vir})}
\sim\frac{M_{\rm halo}}{2M_{\rm disc}}\left(\frac{r_{D}}{r_{s}}\right)\frac{1}{f(c_{\rm vir})}.
\label{fh}
\end{equation}
We introduce the parameters $c'=r_{s}/r_{D}$ and
\begin{equation}
f_{b}=\frac{1}{2}\left(\frac{M_{\rm bulge}}{M_{\rm disc}}\right)\left(\frac{r_{D}}{r_{\rm bulge}}\right).
\label{fb}
\end{equation}
The parameters $f_h$ and $f_b$ measure the importance of the
halo and bulge, respectively, in determining the
dynamics of the flow.
The asymptotic velocity in terms of these parameters is (see equation [\ref{orbit01}])
\begin{equation}
\frac{v_{\infty}^{2}-v_{0}^{2}}{v_{c}^{2}}=\Gamma_{0}\left(1+\frac{f_{h}}{2c'}\right)\frac{\pi}{2}-(1+f_{b}+f_{h}).\label{orbit02}
\end{equation}
Note that for $f_{b},f_{h}\rightarrow0$, equations
(\ref{orbit02}) and (\ref{terminal01}) are equivalent.

We combine the effects of the galactic bulge and dark matter halo
using the parameter $f_{0}=f_{h}+f_{b}$. Assuming $\Gamma_{0}=1$,
the condition for matter to be unbound is
\begin{equation}
f_{0}<\left[\left(\frac{v_{0}}{v_{c}}\right)^{2}+\left(\frac{\pi}{2}-1\right)\right]\left(1-\frac{\pi}{4c'}\right)^{-1}.\label{f0limt}
\end{equation}
More generally, taking $f_{0}$ as a parameter, the terminal velocity
from equation (\ref{orbit02}) is less than the value from equation
(\ref{terminal01}). For the purposes of an estimate, taking
$c'\sim10$, the relation between $v_{\infty}$ and $\langle v_{\rm
rot}\rangle$ becomes
\begin{equation}
v_{\infty}\simeq1.5(1-1.6f_{0})^{1/2}v_{\rm rot},\label{decreas02}
\end{equation}
which shows that the dark matter halo potential well is too deep for
particles to escape for $f_{0}\gtrsim0.6$.

Quantitatively the halo, extended disc, and bulge should have very similar effects in decreasing the outflow velocity. For simplicity, we can first neglect the bulge/disc and focus on $f_{0}=f_{h}$ ($f_{b}=0$). In this case the typical value of $f_{0}$ is given by
\begin{equation}
f_0\sim0.25\left(\frac{M_{\rm halo}/M_{\rm disc}}{30}\right)
\left(\frac{r_D/r_s}{1/30}\right)\left(\frac{2}{f(c_{\rm vir})}\right),
\label{scalef0}
\end{equation}
where we have taken representative values of
$M_{\rm halo}/M_{\rm disc}\sim10-100$ (e.g., \citealt{Leauthaud11}),
$c^\prime=r_s/r_D\sim10-100$, and $f(c_{\rm vir})\sim2$
(for $c_{\rm vir}\sim15$; \citealt{Maccio08}).
Thus, instead of the naive estimate without
the dark matter halo in equation (\ref{velocity_1}), which
yields $v_\infty\simeq1.5\,v_{\rm rot}$, we obtain
$v_\infty\simeq v_{\rm rot}$.
Even so, equation (\ref{decreas02}) makes it clear that
the relation $v_\infty\simeq1.5\,v_{\rm rot}$
from equation (\ref{velocity_1}) cannot be strictly achieved for Eddington-limited rapidly star-forming galaxies
by the mechanism discussed in this paper since all
systems have dark matter haloes,
not to mention the fact that Oppenheimer \& Dav\'{e}'s model includes an additional kick
to increase the wind terminal velocity.
To the extent that the ``2'' in the expression of Oppenheimer \& Dav\'{e}'s model $v_{\rm \infty}\sim2\,v_{\rm rot}$ or ``$\sim3-5$'' with an extra kick is required to match observations and simulations, additional physics such as an effectively super-Eddington galaxy
luminosity or supernova-driven hot flows would be necessary to
add to the models here (e.g., \citealt{Murray11}; \citealt{Hopkins11a,Hopkins11b}).

\begin{figure}
\centerline{\includegraphics[width=9cm]{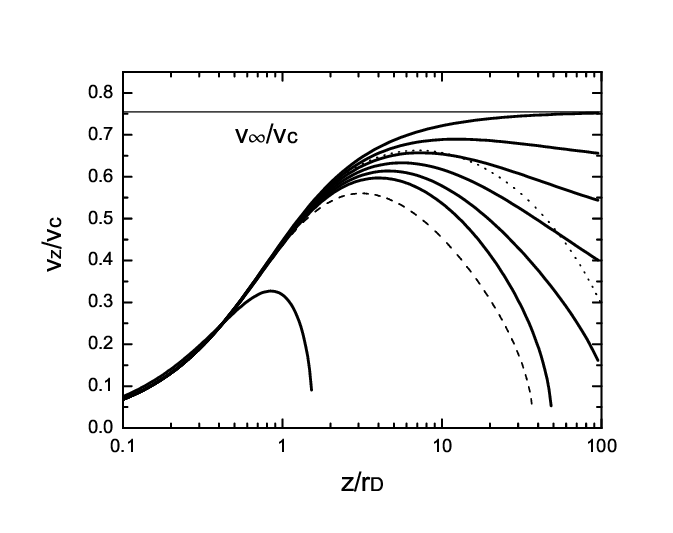}}
\caption{Outflow velocity $v_{z}/v_{c}$ along the pole with $c'=10$ and $f_{0}$
from top down $f_{0}=0,0.2,0.4,0.6,0.8,1,5$ (thick lines), $c'=4$ (dashed line) and $c'=100$
(dotted line) with $f_{0}=1$. The horizon line labeled as $v_{\infty}/v_{c}$ is the analytical solution
as in equation (\ref{terminal01}).}\label{fig_polevelocity}
\end{figure}

The quantitative scaling relations for bound and unbound
outflows, encapsulated in the factors $f_0$ and $c'$, are illustrated
in Figure \ref{fig_halo}. The left panel shows contours of the
asymptotic kinetic energy $(v_{\infty}/v_{c})^{2}$ as a function of
$f_{0}$ and $c'=r_s/r_D$ assuming $v_{0}=0$ for unbound
outflows.\footnote{Increasing $v_{0}$ to $0.2v_{c}$ does not change
the position of the contours appreciably.} The right panel of Figure
\ref{fig_halo} shows contours of $\log_{10}[z_{\rm turn}/r_D]$, the
turning point of the wind scaled to the disc radius for bound
``outflows.'' According to equation (\ref{scalef0}), if all else being equal, a disc with higher ratio of $M_{\rm halo}/M_{\rm disc}$
has larger $f_{0}$, smaller $z_{\rm turn}$, and the flow has a shorter timescale
 for reaching the turning point $t_{\rm turn}\sim z_{\rm turn}/v_{c}\sim t_{c}(z_{\rm
turn}/r_{D}$) (see equation [\ref{tc}]). The right panel of Figure
\ref{fig_halo} shows that $z_{\rm turn}/r_D$ can be larger than $\sim1000$,
implying that the flow reaches many tens of kpc in
height above the disc before falling back towards the host on a
timescale of $\sim$\,Gyr.

Importantly, even if $f_0$ is large enough that the flow is
bound,  one may still observe an outgoing wind from a rapidly star-forming
galaxy for two reasons.
First, there is a region of parameter space where the time to
reach the turning point $t_{\rm turn}$ is larger than the time
for the rapidly star-forming galaxy to deplete its gas supply $t_\star=M_g/{\rm SFR}$ (see
equation [\ref{tstar}]).  Since $t_{\rm turn}/t_\star\sim(z_{\rm turn}/r_D)
t_c/t_\star$ (see equation [\ref{tc}]), $z_{\rm turn}/r_D\gtrsim60r_{D,\,1kpc}^{-1/2}
f_{g,0.5}^{-0.4}\Sigma_0^{0.1}$ is required for $t_{\rm turn}/t_\star\gtrsim1$.
Second, an observable flow along the line of
sight to a non-edge-on disc can have a maximum velocity $v_{\rm max}$
that is comparable to $v_{\infty}$ in equation (\ref{terminal01}),
even though the flow is bound on large scales ($z_{\rm turn}$)
by the dark matter halo. Figure \ref{fig_polevelocity} shows the one-dimensional outflow
velocity $v_{z}/v_{c}$ along the pole with various $f_{0}$ and $c'$.
In the bound case, the maximum outflow velocities $v_{\rm max}$
peak around $1-10r_{D}$ before decreasing to much lower values.
The maximum $v_{\rm max}$ is still comparable to $v_{\infty}$
in the absence of a halo or bulge or extended disc, except for very
large $f_{0}\gtrsim5$ (see equation [\ref{scalef0}]).   The
velocity of the flow before changing sign and falling back to the
host galaxy is time-dependent and approximately reaches its maximum
at a time $\sim t_{c}$ (equation [\ref{tc}]).

The fact that $t_{\rm turn}>t_\star$ and that $v_{\rm max}$
approaches a few $\times v_{\rm rot}$ together imply that
if galaxies are selected as bright
rapidly star-forming galaxies, they may appear to have unbound outflows even though
the gas is in fact bound on large scales by the dark matter halo potential.
Indeed, Figure \ref{fig_polevelocity} implies that
rapidly star-forming galaxies with bound flows may exhibit a rough
correlation of the form  $v_{\rm max}\sim1-1.5\,v_{\rm rot}$.

\begin{figure*}
\begin{center}
\centerline{
\includegraphics[width=18cm]{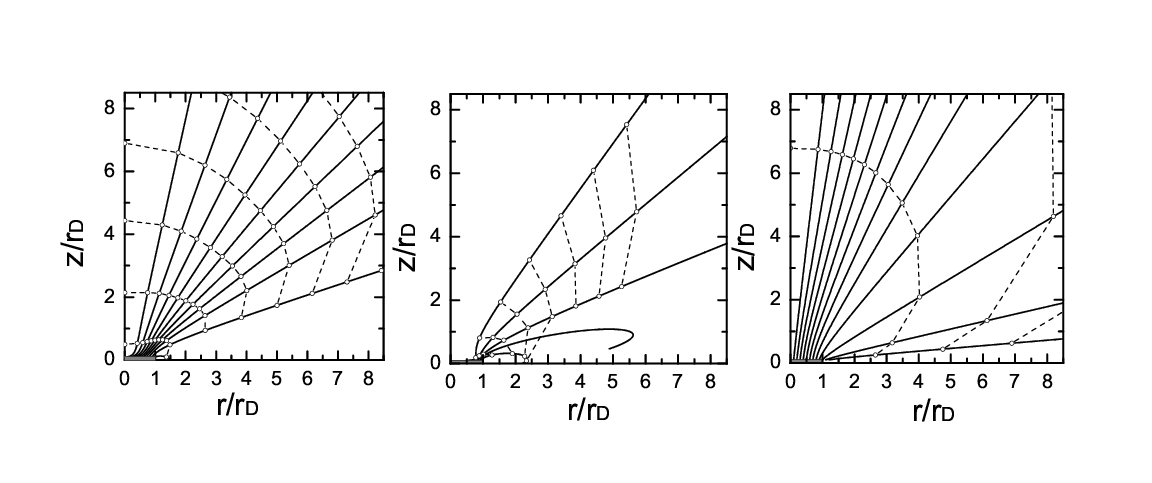}}
\vspace*{-2cm} \centerline{\includegraphics[width=18cm]{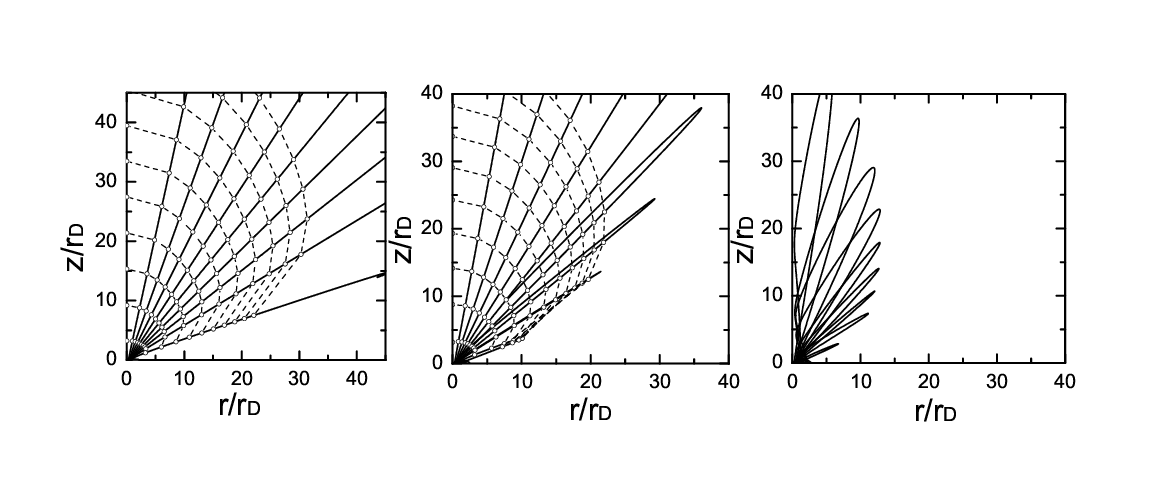}}
\vspace*{-.5cm} \caption{The particle orbits from the disc without
(upper panels) and with (lower panels) NFW potential. Solid lines
show 2-dimensional projections of the 3-dimensional orbits. Upper
panels, left to right, we take $\Gamma_{0}=1,0.88$ and 5.
Particles are initially located at disc surface $z_{0}=0.1r_{D}$ with $r_{0}/r_{D}=0$, 0.1, 0.2, 0.3, 0.4, ... 1.0. In the upper middle panel trajectories with initial radii $r_{0}/r_{D}< 0.5$ are skipped because they are bound, while in the upper right panel extra trajectories with initial radii $r_{0}/r_{D}=1.1$ and 1.2 are also plotted. The particle positions and constant time surfaces are labeled at
$t/t_{c}=2$, 4, 6, 8, 10..., where $t_{c}=\sqrt{r_{D}/(4G\Sigma)}$
(eq.~\ref{tc}). Lower panels, left to right: $f_{0}=0.2$, 0.6, 1.0
with $\Gamma_{0}=1$ and $c'=10$. Constant time surfaces are marked
at $t/t_{c}$=5, 10, 15, 20, 25....}\label{fig_orbit_1}
\end{center}
\end{figure*}
\begin{figure*}
\centerline{\includegraphics[width=18cm]{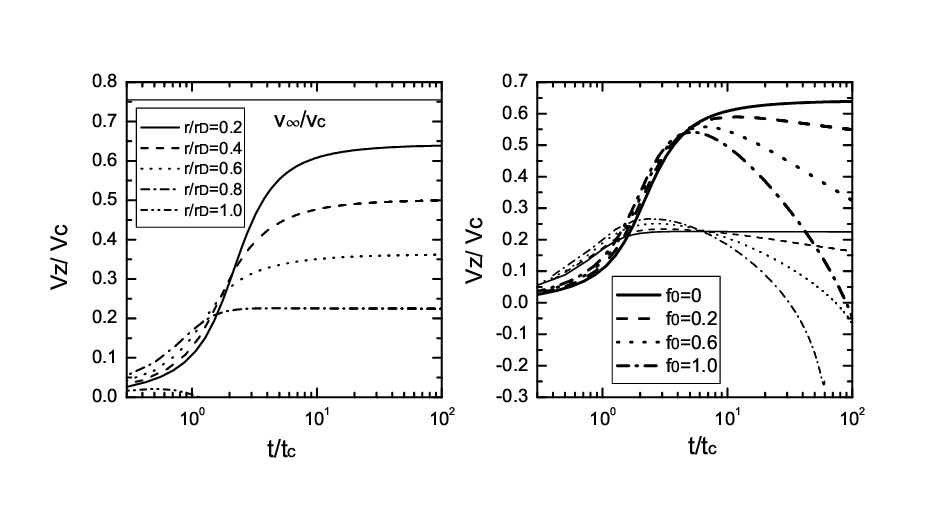}} \vspace*{-.5cm}
\caption{\textit{Left panel}: The time evolution of $v_{z}/v_{c}$
for $r/r_{D}=0.2$ to 1.0 starting with $z/r_{D}=0.1$ and
$\Gamma_{0}=1$. The horizon line $v_{\infty}/v_{c}$ is the same as
Figure \ref{fig_polevelocity}. \textit{Right panel}: the velocity
$v_{z}/v_{c}$ starting from $(r/r_{D},z/z_{D})=(0.2, 0.1)$ (thick
lines) and $(0.8,0.1)$ (thin lines) with different values $f_{0}$
for the dark matter potential with $c'=10$ and
$\Gamma_{0}=1$.}\label{fig_velocity_1}
\end{figure*}

\section{3-Dimensional Winds and Mass Loss Rate}

So far we have focused on the forces along the polar $z$-axis.
Figure \ref{fig_orbit_1} shows 2-dimensional (2D) projections of the
3D orbits of test particles accelerated by radiation pressure and
gravity, as well as their constant time surfaces above a uniform
disc, starting from an initial height $z_{0}=0.1r_{D}$ and initial radii $r_{0}=0,0.1r_{D},0.2r_{D},0.3r_{D}$..., both with and
without an NFW potential, and with no galactic bulge or extended passive disc. The opening angles of particle trajectories are formed because
particles initially co-rotate in the disc with an angular omentum $v_{\rm rot}^{2}/r$ except for particles along the $z$-axis. In the
absence of a halo, the upper middle panel shows that winds can be
driven from part of the disc region (e.g., $r_{0}>0.5r_{D}$ in this panel) even when the disc is sub-Eddington ($\Gamma_{0}=0.88$),
while the upper right panel shows the wind can also be launched even beyond the edge of bright disk ($r>r_{D}$) in a super-Eddington case ($\Gamma_{0}=5$).
The lower panels show that the character of the flow changes significantly
as $f_{0}$ increases from 0.2, to 0.6, to 1.0 (left to right; compare with
equation [\ref{scalef0}]). For the case $f_{0}=0.2$, the wind is clearly unbound.
For $f_{0}=0.6$, particles emerging near the
$z$-axis are accelerated to very large vertical distances, whereas
particles emerging from the outer disc region fall back to the
disc rapidly.  A more massive halo with
$f_{0}=1.0$ produces only a ``fountain flow'' in which particles fall
back to the disc with a timescale of $\sim0.1-1$ Gyr. Moreover,
Figure \ref{fig_velocity_1} gives the 3D orbit vertical component
$v_{z}$ evolution of the particles driven from different disc
regions. For a face-on disc, the velocity $v_{z}$ of the wind is
just the velocity along the line of sight. The maximum velocities
are reached roughly at $t\sim t_{c}$. Also, the maximum and terminal
velocities of particles from the outer disc region are smaller than
those from the inner disc region. As in Figure \ref{fig_orbit_1},
matter from the outer region of the disc can be bound by the halo's
gravitational potential even though matter from the inner region is
unbound.

To estimate the rate of mass ejection and mass loss rate from Eddington-limited discs, we
first calculate that the Eddington flux and luminosity are
\begin{eqnarray}
F_{\rm Edd}=2\pi cG\Sigma/\kappa\sim3\times10^{12}\Sigma_{0}\kappa_{1}^{-1}\;L_{\odot}\;\textrm{kpc}^{-2},
\label{fedd}
\end{eqnarray}
and
\begin{equation}
L_{\rm Edd}=\pi r_{\rm rad}^{2}F_{\rm Edd}\sim10^{13}\left(\frac{r_{\rm rad}}{r_{D}}\right)^{2}\Sigma_{0}\kappa_{1}^{-1}r_{D,1kpc}^{2}\;L_{\odot},
\end{equation}
respectively.

The total mass ejection rate from the disc surface $\dot{M}_{\rm ej}$
is a local quantity measured on a disc scale height $h_{\rm disc}$.
The disc, bulge, and dark matter halo gravitational forces at the disc scale height $h_{\rm disc}$ estimated at $(R,z)\sim(0,h_{\rm disc})$ as $f_{\rm disc}\sim 2\pi G \Sigma_{(\rm disc)}$, $f_{\rm bulge}\sim\pi G \rho_{\rm bulge}h_{\rm disc}$, and $f_{\rm halo}\sim GM_{\rm halo}/[2r_{s}^{2}f(c_{\rm vir})]$ respectively. Taking the case of $\rho_{\rm bulge}h_{\rm disc}\ll\Sigma_{\rm disc}$, the gravity of the bulge can be neglected at $h_{\rm disc}$ since $f_{\rm bulge}\ll f_{\rm disc}$. Similarly, since the ratio of the halo to the disc gravity $f_{\rm halo}/f_{\rm disc}\sim4\times10^{-3}\left(\frac{r_{D}/r_{s}}{1/30}\right)^{2}\left(\frac{M_{\rm halo}/M_{\rm disc}}{30}\right)\left(\frac{2}{f(c_{\rm vir})}\right)\ll1$, the halo can also be neglected on the disc height scale $h_{\rm disc}$.
Therefore $\dot{M}_{\rm ej}$ can be determined only by the disc, and be estimated
in the absence of bulge and halo from the integrated momentum equation.
In the single-scattering limit (i.e., all photons are
scattered/absorbed once in the wind; e.g., MQT05),
an estimate of $\dot{M}_{\rm ej}$ is
\begin{equation}
\dot{M}_{\rm ej}v_{\infty,f_{0}=0}\sim L_{\rm Edd}/c,\label{massloss01}
\end{equation}
where $v_{\infty,f_{0}=0}$ is the terminal velocity of the flow without
a bulge, extended passive disc or dark matter halo (equation [\ref{terminal01}]).
Since $v_{\infty,f_{0}=0}$ is an upper limit to the velocity of the
flow (valid as $f_0\rightarrow0$; see equation [\ref{decreas02}]),
equation (\ref{massloss01}) is only approximate.  Nevertheless,
to the extent that $v_{\rm max}$ is of order $v_{\infty,f_{0}=0}$
(see Figure \ref{fig_polevelocity}), equation (\ref{massloss01})
should yield an order-of-magnitude estimate of the mass loss
rate from the disc itself.
Combining equations (\ref{terminal01}) and (\ref{massloss01}) we have
\begin{equation}
\dot{M}_{\rm ej}\sim3\times10^{2}\left(\frac{r_{\rm rad}}{r_{D}}\right)^{2}\Sigma_{0}^{1/2}\kappa_{1}^{-1}r_{D,1kpc}^{3/2}
\;M_{\odot}\;\textrm{yr}^{-1}\label{massloss02},
\end{equation}
which is similar to observational results (e.g., Martin 2005, 2006).
Using equations (\ref{terminal02}) and (\ref{Fedd}), the mass ejection rate is also given by
\begin{equation}
\dot{M}_{\rm ej}\sim6\times10^{2}\;f_{g,0.5}^{-0.5}\kappa_{1}^{-1}r_{D,1kpc}^{0.79}
\left(\frac{\rm SFR}{50M_{\odot}\textrm{yr}^{-1}}\right)^{0.36}\;M_{\odot}\;\textrm{yr}^{-1}\label{massloss02b}.
\end{equation}
Using $L=\epsilon c^2\, {\rm
SFR}$ to calculate the luminosity of a rapidly star-forming galaxy (e.g., Kennicutt 1998),
where $\epsilon_{-3}=\epsilon/10^{-3}$ is the efficiency
with which star formation converts mass into radiation,
the single-scattering estimate for the ratio of the mass
ejection rate to the SFR is
\begin{equation}
\frac{\dot{M}_{\rm ej}}{\rm SFR}\sim\frac{\epsilon c}{v_{\infty,f_0=0}}
\sim0.8\,\epsilon_{-3}\Sigma_0^{-1/2}r_{D,\,1kpc}^{-1/2},\label{massloss03}
\end{equation}
which implies that radiation pressure drives more matter from discs
in low-mass galaxies (see MQT05): for example, $\dot{M}_{\rm ej}/{\rm
SFR}\gtrsim10$ for $\Sigma r_D\lesssim3.1\times10^4$\,M$_\odot$
pc$^{-1}$.
However, as low mass galaxies are usually the most dark
matter halo dominated with higher ratio of $M_{\rm halo}/M_{\rm disc}$(\citealt{Persic96}), or larger $f_0$
(see equations [\ref{f0limt}] and [\ref{scalef0}]), the matter lost from
low mass discs may not escape the halo potential, and will fall
back on the timescale $t_{\rm turn}$ (see the right panel of Figure \ref{fig_halo}).
Otherwise, some additional physical mechanism beyond that described in
this paper (e.g., a supernova-heated wind) is needed to unbind it from
the halo.

The quantity $\dot{M}_{\rm ej}$ in the above expressions is
the rate at which matter is ejected from the disc in either  a
bound or unbound flow.
In the former case, as discussed in Section 3, the outflow reaches a maximum outward velocity
$v_{\rm max}$ before returning to the disc on a timescale $t_{\rm turn}$,
which in many cases is $>t_\star$. In the case of an unbound outflow,
$\dot{M}_{\rm ej}$ is an estimate for the mass loss rate from
the halo as a whole on large scales. As emphasized in Section 3, the critical
parameter determining whether or not the flow is bound or unbound
is $f_0$ (equations [\ref{f0limt}] and [\ref{scalef0}]).

\begin{figure}
\centerline{\includegraphics[width=9cm]{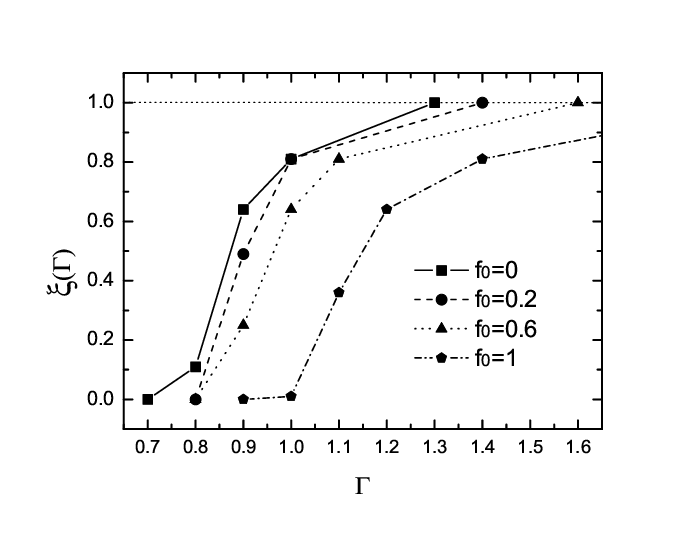}}
\caption{Ratio of the asymptotic mass loss rate from the gravitational potential($\dot{M}_{\infty}$) and the initial mass loss rate from the disc ($\dot{M}_{\rm ej})$, $\xi(\Gamma)=\dot{M}_{\infty}/\dot{M}_{\rm ej}$, as the function of disc Eddington ratio $\Gamma$, assuming a NFW dark matter potential as in Figure \ref{fig_orbit_1} with $f_{0}=0,0.2,0.6$ and 1 with $c'=10$.}\label{fig_ratio}
\end{figure}
For a totally unbound outflow that escapes the gravitational potential, the asymptotic mass loss rate of the outflow from the large-scale bulge and halo gravitational potential ($\dot{M}_{\infty}$) should be the same as the mass loss rate from the disc $\dot{M}_{\rm ej}$: $\dot{M}_{\infty}=\dot{M}_{\rm ej}$. However, in the bound and partially bound cases (e.g., see Figure \ref{fig_orbit_1} as examples), $\dot{M}_{\infty}$ should only be a fraction of $\dot{M}_{\rm ej}$ such that $\dot{M}_{\rm ej}>\dot{M}_{\infty}\geq0$, and the ratio $\xi=\dot{M}_{\infty}/\dot{M}_{\rm ej}$ depends on the disc Eddington ratio $\Gamma$ and the depth of the large scale gravitational potential. Figure \ref{fig_ratio} shows $\xi(\Gamma)$ for NFW haloes with $c'=10$ and $f_{0}=0,0.2,0.6$ and 1. We calculate $\xi(\Gamma)$ by following the trajectories of test particles, as in Figure \ref{fig_orbit_1}, launched from various radii just above the disc surface. Because the disc is assumed to have uniform surface density, $\xi(\Gamma)$ is the ratio of the disc area over which particles terminally escape the large scale halo to the total disc area. We find that $\xi(\Gamma)$ is a strongly increasing function of $\Gamma$, and that it decreases substantially as $f_{0}$ is increased. For $f_{0}=0.2$ and 0.6 (see equation [\ref{scalef0}]), $\xi(\Gamma)\simeq0.8$ and 0.65 at $\Gamma=1$, respectively, showing that for typical dark matter halo properties $\sim60\%-80\%$ of the matter ejected from the disc is completely unbound for Eddington discs, whereas the remainder forms a fountain flow. The totally unbound wind (i.e., $\xi=1$) can be only approached for super-Eddington discs. A more complete study of the asymptotic mass loss should follow the hydrodynamics of the outflow instead of taking the test particle approximation.

\section{Conclusions \& Discussion}
In this paper we study the large-scale winds from uniformly bright
self-gravitating discs radiating near the Eddington limit. Different
from the spherical case, where the Eddington ratio $\Gamma=f_{\rm
rad}/f_{\rm grav}$ is a constant with distance from the source, for
discs $\Gamma$ increases to a maximum of twice its value at the disc
surface (Section 2). As a result, such discs radiating at (or even
somewhat below; see Figures \ref{fig_subEdd} and \ref{fig_orbit_1}) the Eddington limit are
unstable to driving large-scale winds by radiation pressure.

We quantify the characteristics of the resulting outflow in the
context of Eddington-limited star-forming galaxies, motivated by the work of TQM05
who argued that radiation pressure on dust is the dominant feedback
process in rapidly star-forming galaxies. We find that the asymptotic terminal
velocity along the polar direction from discs without a stellar
bulge, extended passive disc or dark matter halo is $v_{\infty}\sim\sqrt{4\pi G\Sigma
r_{D}}\sim1.5\,v_{\rm rot}$, where $r_{D}$ is the disc
radius, $\Sigma$ is the surface density, $v_{\rm rot}$ is the disc
rotation velocity (see equations [\ref{terminal01}] and [\ref{velocity_1}]),
and may range from $\sim50-1000$\,km s$^{-1}$ for rapidly star-forming galaxies,
depending on the system considered. Furthermore, by employing the
observed Schmidt law, we find that $v_{\infty}\propto$
SFR$^{0.36}$r$_{D}^{-0.21}$ (see equation [\ref{terminal02}]). These
results, in the absence of dark matter halo and bulge, or extended passive disc, are in
agreement with recent observations (e.g., \citealt{Martin05,Martin06}; \citealt{Weiner09}; see also \citealt{Chen10}). The typical mass loss rate
from an Eddington-limited disc in the single-scattering limit is
given by equation (\ref{massloss02}) and suggests these outflows may
efficiently remove mass from the disc (equation [\ref{massloss03}]).

However, both wind velocities and outflow rates can be
significantly decreased by the presence of a spherical old stellar
bulge or extended gravitational disc or dark matter halo potential (Section 3).
Deeper or more extended spherical gravitational potentials cause the flow to be bound on large scales, and to
produce only ``fountain flows'' where particles fall back to the disc on a
typical timescale of $\sim0.1-1$ Gyr, depending on the parameters of the
system considered (see
Figures \ref{fig_halo}-\ref{fig_ratio}). The criterion for the
flow to become bound along the polar direction is given in equation (\ref{f0limt}).
For typical values of the parameter $f_0$ (see equation [\ref{scalef0}]),
we find that the winds from rapidly star-forming galaxies can be either bound or unbound
(see Figure \ref{fig_halo}). As an example, for $f_{0}\simeq0.25$ in equation (\ref{scalef0}),
the asymptotic velocity of the wind is decreased by a factor of $\simeq0.7$
from the case neglecting the dark matter halo completely (compare equations [\ref{velocity_1}]
and [\ref{decreas02}]), from $\simeq1.5\,v_{\rm rot}$ to
$\simeq v_{\rm rot}$. However, for $f_{0}>0.6$ the asymptotic velocity goes to zero along the $z$-xis and the flow becomes bound on large scales.
A more elaborate 3D calculation shows that a totally unbound wind from a 2D disc surface can only be approached at super-Eddington discs, otherwise the outflow asymptotic mass loss rate from the large scale gravitational potential $\dot{M}_{\infty}$ is only a fraction of mass ejection rate from the disc $\dot{M}_{\rm ej}$ (Figure \ref{fig_ratio}).

Importantly, even in the limit of bound fountain flows, if the timescale $t_{\rm turn}$
for reaching the turning point $z_{\rm turn}$ is longer than the
lifetime of the rapidly star-forming galaxy $t_{\star}$ (equation [\ref{tstar}]), one may
still observe an outward going wind while the rapidly star-forming galaxy is active
and bright. The maximum positive velocity of the flow $v_{\rm max}$
along the line of sight for non-edge-on discs is still correlated
with $v_{\rm rot}$ of the disc (see Section 3). These facts may complicate the
inference from observations of winds that they are unbound
from the surrounding large-scale dark matter halo.

More detailed work is required to fully assess radiation
pressure on dust as the mechanism for launching cool gas from
rapidly star-forming discs. Extra observational evidence beyond the relation
$v_{\infty}\propto$ SFR$^{0.36}$ should certainly be added to verify the
radiation-driven model from more realistic discs instead of the idealized model of uniform discs we consider in this paper.
Equations (\ref{terminal01}) and (\ref{terminal02}) also
give testable relations between $v_{\infty}$ and galactic surface density $\Sigma$, which
can also be used to compare with observations of cold winds
from galaxies with observed radii and masses.
One key ingredient of our model is that galaxies should approach the
dust Eddington limit. TQM05 proposed that a significant fraction of the radiation from ULIRGs should be
produced by an Eddington-limited rapidly star-forming galaxy
(see also the discussion of the dust Eddington limit in \citealt{Hopkins10}).
If so, the dust-driven mechanism
discussed in this paper can be applied to the winds from ULIRG discs.
Moreover, galaxies at high redshifts are more active
and thus potentially more likely to approach the Eddington limit (e.g., MQT05).
Take the data on extreme starburst Arp220 in \cite{1998ApJ...507..615D} as a case study.
The surface density of Arp 220 West (East) core region is $\Sigma\approx20$ g cm$^{-2}$ (18 g cm$^{-2}$),
corresponding to a FIR-thick disc with an Eddington ratio
of $\Gamma\approx0.3 (\kappa/10$ cm$^{2}$ g$^{-1})$ ($\Gamma\approx0.1 (\kappa/10$ cm$^{2}$ g$^{-1})$).
Ignoring the observational uncertainties in $\Gamma$,
the sub-Eddington luminosities imply that the outflow from the Arp 220 cores may be
caused by other mechanisms such as the energy and
momentum input from supernovae rather than just radiation pressure on dust
(\citealt{CC85}; \citealt{SH09}; \citealt{Hopkins11a, Hopkins11b}).

The main limitations of our model of radiation-driven wind presented here are
that (1) we consider only discs with constant brightness and surface
density, (2) we do not compute the hydrodynamics of the flow, but
instead treat the outflow in the test-particle limit, and (3)
we ignore other physical mechanisms,
which may act in concert with radiation pressure (MQT05; \citealt{Murray11}).
The general case, with hydrodynamics, realistic disc brightness, surface
density, and dust opacity profiles will modify the picture presented
here. Such an effort is underway. However, the basic conclusion that uniformly bright
self-gravitating discs radiating near the Eddington limit are able
to drive large-scale winds --- particularly in the high-$\Sigma$
limit in rapidly star-forming galaxies (see Section 2)--- should not be fundamentally changed
by more elaborate considerations. Indeed, although we have
specialized the discussion to rapidly star-forming galaxies and dust opacity, the
instability derived in Section 2 is of general applicability.

\section*{Acknowledgments}
We thank the anonymous referee for his/her very useful
comments that have allowed us to improve our paper. We also thank Norman Murray, Crystal Martin, Romeel
Dav\'{e}, Paul Martin, and especially Mark Krumholz and Eliot Quataert for many
stimulating discussions and for a critical reading of the text. This
work is supported by NASA grant \# NNX10AD01G.

\end{document}